\documentclass[11pt,a4paper,sans-serif]{article} 
\usepackage{graphicx,epsfig} 
\usepackage[top=1in,bottom=1in,left=1in,right=1in]{geometry} 
\usepackage{color} 
\usepackage[usenames,dvipsnames,svgnames,table]{xcolor} 
\usepackage{colortbl} 
\usepackage{rotating} 
\usepackage{amssymb,amsmath,mathtools} 
\usepackage{footnote} 
\usepackage{booktabs} 
\usepackage{tabularx}
\usepackage{multicol}
\usepackage{float}

\title{Complexions in a 
modified Langmuir-McLean model of
grain boundary segregation}
\author{S.R. Wilson}
\date{}

\def\lejcek{1}
\def\suttonballuffi{2}
\def\kirchheim{3}
\def\koch{4}
\def\gibbs{5}
\def\deSilva{6}
\def\kuzmina{7}
\def\raabeherbig{8}
\def\huber{9}
\def\mclean{10}
\def\langmuir{11}
\def\fowlerguggenheim{12}
\def\seahhondros{13}
\def\wynblattchatain{14}

\def\citeTab{\hspace{5pt}}
\def\wynblattchatainCitation{
[\wynblattchatain]\citeTab
P Wynblatt and D Chatain. ``Anisotropy of segregation at
grain boundaries and surfaces.''
Metall Mater Trans A 37 (2006) 2595-2620
}
\def\kirchheimCitation{
[\kirchheim]\citeTab
R Kirchheim. ``Reducing grain boundary, dislocation line
and vacancy formation energies by solute segregation. I.
Theoretical background.'' Acta Mater. 55 (2007) 5129-5128
}
\def\kochCitation{
[\koch]\citeTab
C Koch, R Scattergood, K Darling, and J Semones.
``Stabilization of nanocrystalline grain sizes by solute
addition.'' J Mater. Sci. 43 (2008) 7264-7272
}

\def\raabeherbigCitation{
[\raabeherbig]\citeTab
D. Raabe, M. Herbig, S. Sandl\"obes, Y.Li, D. Tytko,
M. Kuzmina, D. Ponge, and P.P. Choi.
``Grain boundary segregation engineering in metallic alloys:
A pathway to the design of interfaces.''
Curr Op Solid St and Mat Sci 18 (2014) 253-261
}
\def\huberCitation{
[\huber]\citeTab
L. Huber, R. Hadian, R. Grabowski, and J. Neubeauer.
``A machine learning approach to model solute 
grain boundary segregation.'' npj Comput Mater 4, 64 (2018)
}

\def\mcleanCitation{
[\mclean]\citeTab
D. McLean, \textit{Grain Boundaries in Metals}.
Clarendon Press (1957)
}
\def\langmuirCitation{
[\langmuir]\citeTab
I. Langmuir, ``The adsorption of gases on plane
surfaces of glass, mica, and platinum.''
J. Am. Chem. Soc., 40(9), 1361-1403 (1918)
}
\def\seahhondrosCitation{
[\seahhondros]\citeTab
MP Seah and ED Hondros,
``Grain boundary segregation."
Proc. Roy. Soc. Lond. A 335, 191 (1973)
}
\def\gibbsCitation{
[\gibbs]\citeTab
JW Gibbs, \textit{The Collected Works of J. Willard Gibbs, Ph.D, LLD}.
Yale University Press, London, England, Vol. 1-2 (1957)
}
\def\fowlerguggenheimCitation{
[\fowlerguggenheim]\citeTab
RH Fowler and EA Guggenheim,
\textit{Statistical thermodynamics.}
Cambridge University Press (1939)
}
\def\deSilvaCitation{
[\deSilva]\citeTab
AK de Silva, RD Kamachali,
D Ponge, B Gault,
J Neugebauer, and D Raabe,
``Thermodynamics of grain boundary segregation,
interfacial spinodal and their relevance for
nucleation during solid-solid phase transitions.''
Acta Mater 168, 109-120 (2019)
}
\def\lejcekCitation{
[\lejcek]\citeTab
P Lejcek, \textit{Grain Boundary Segregation in Metals}.
Springer, Science and Business Media (2010)
}
\def\suttonballuffiCitation{
[\suttonballuffi]\citeTab
AP Sutton and RW Balluffi. \textit{Interfaces in 
Crystalline Materials.} Clarendon Press (1995)
}
\def\kuzminaCitation{
[\kuzmina]\citeTab
M Kuzmina, D Ponge, and D Raabe.
``Grain boundary segregation engineering
and austenite reversion turn embrittlement
into toughness:
Example of a 9 wt. \% medium Mn steel.''
Acta Mater. 86, 182 (2015)
}

\def\citations{
\lejcekCitation\\
\suttonballuffiCitation\\
\kirchheimCitation\\
\kochCitation\\
\gibbsCitation\\
\deSilvaCitation\\
\kuzminaCitation\\
\raabeherbigCitation\\
\huberCitation\\
\mcleanCitation\\
\langmuirCitation\\
\fowlerguggenheimCitation\\
\seahhondrosCitation\\
\wynblattchatainCitation\\
}

\begin{document}
\maketitle
\fontfamily{cmr}
\selectfont
\begin{abstract}
\noindent
The Langmuir-McLean isotherm is often
interpreted as providing an approximation to the 
most probable grain boundary segregation
as a function of the bulk mole 
solute fraction $x_B$,
even though $x_B$ is not an independant parameter
in the free energy minimization on which
it is based. In this paper it is shown that 
the most probable segregation for a specified $x_B$ 
differs from the standard Langmuir-McLean relation.
Numerical solution of the derived equation suggests
that two potentially stable interface compositions
are associated with most bulk compositions.
One solution represents a state with
an excess of solute along
the boundary relative to the bulk, 
while the other represents a deficit.
The vacancy content ratio between the interface and the bulk
plays a large role in determining the shape of the derived isotherm.
\end{abstract}

\begin{multicols}{2}
\section{Introduction}
Segregation is the process
of grouping impurities
and structural defects together in a material
system
[\lejcek,\suttonballuffi].
Any free energy reduction 
associated with segregation
can be leveraged to 
stabilize a desired defect 
structure, allowing 
materials engineers to ``bake in''
what would otherwise be transitory
material properties that depend
on the dominant defect population.
[\kirchheim,\koch]

Segregation has therefore
been the subject of 
much research
in metallurgy and materials science
from its initial roots [\gibbs]
to the present day [\deSilva,\kuzmina,\raabeherbig,\huber].
The simplest model of 
equilibrium segregation 
along a grain boundary is
given by the Langmuir-McLean isotherm
[\mclean,\langmuir]
\begin{align}
\frac{\Gamma_B}{\Gamma_0-\Gamma_B}
=\frac{x_B}{x_A}\text{e}^{-\delta G/k_BT}
\label{mcleanisotherm}
\end{align}
where 
$\delta G$
is the segregation free energy per solute atom,
$k_B$ is Boltzmann's constant,
$T$ is the ambient temperature,
$\Gamma_B$ is the number density of solute atoms
segregated to the boundary
with maximal value $\Gamma_0$,
and
$x_B$ and $x_A$ are the mole fractions
of components A and B in the bulk.

The Langmuir-McLean isotherm serves as a
common touchpoint for a number of segregation models
proposed over the years
on the basis of more complicated assumptions.
A few of the more prominent models include
the Fowler-Guggenheim isotherm [\fowlerguggenheim],
which considers the influence of solute-solute 
interactions in the interface, and
the Seah-Hondros isotherm [\seahhondros], 
derived on the basis of solid-state theoretic methods.

Despite these efforts, there remains a significant
discrepency between the observed and predicted 
segregation to interfaces in many real materials [\wynblattchatain].
In this paper I present a simple modification 
to the Langmuir-McLean model that yields very different
predictions.

\section{Model derivation}

Equation (\ref{mcleanisotherm}) results from 
analyzing a two-state model of a grain boundary
in which impurity atoms of component B are either segregated
to the interface or are free to roam in a bulk matrix
consisting of atoms of component A.
The same two-state system will be considered in this work.
A brief outline of the steps involved
in the derivation of equation (\ref{mcleanisotherm})
will be presented before indicating the changes proposed
in this paper.

Distribute $n_A=n_{A0}+n_{A1}$ atoms of component A
and $n_B=n_{B0}+n_{B1}$ atoms of component B
among $N_1$ indistinguishable bulk sites 
and $N_0$ indistinguishable interface sites,
such that $n_{B0}$ and $n_{A0}$ are
situated on the interface while the rest remain
in the bulk. 
The quantity of primary interest is $n_{B0}$,
because this represents the number of atoms of component B
segregated to the interface. Subscript 1 will indicate
bulk quantities; subscript 0 will indicate interface quantities.

The most probable number
of segregated solute atoms, $n_{B0}$,
minimizes the Helmholtz free energy $F=U-TS$
in the NVT ensemble.
To find this minimum,
we can express
both the internal energy $U$
and the entropy $S$ as functions of $n_{B0}$ and
evaluate $\partial F/\partial n_{B0}=0$. 
An expression for the entropy $S=k_B\ln\Omega$
may be determined by counting the total number 
$\Omega$ 
of indistinguishable
configurations of the system that correspond to 
a specified system configuration
$\lbrace n_{A0},n_{A1},n_{B0},n_{B1}\rbrace$.
Basic combinatorics yields 
\begin{align}
\Omega=
\frac{N_1!}{n_{B1}!n_{A1}!n_{V1}!}\frac{N_0!}{n_{A0}!n_{B0}!n_{V0}!}.
\label{mcleanomega}
\end{align}
where 
$n_{V1}=N_1-n_{A1}-n_{B1}$ 
and
$n_{V0}=N_0-n_{A0}-n_{B0}$ 
are the number of vacant
sites in the bulk and in the interface.
Using this expression, it can be shown that 
the general solution to 
$\partial F/\partial n_{B0} = 0$ 
in the Stirling approximation satisfies
\begin{align}
n_{B0}^{}
n_{A0}^{n_{A0}'}
n_{V0}^{n_{V0}'}
=
n_{B1}^{-n_{B1}'}
n_{A1}^{-n_{A1}'}
n_{V1}^{-n_{V1}'}
\text{e}^{-\delta G/k_BT}
\label{generalisotherm}
\end{align}
where $\delta G=\partial U/\partial n_{B0}$ is the segregation
free energy and primes indicate differentiation with respect to 
$n_{B0}$. 
In order to evaluate the primed exponents that appear
in equation (\ref{generalisotherm}), it is
necessary to specify how each variable depends on 
$n_{B0}$. The Langmuir-McLean isotherm follows from 
imposing the constraints
\begin{align}
n_{A0}+n_{A1}&=n_A \label{lmconstraint1}\\
n_{B0}+n_{B1}&=n_B \label{lmconstraint2}\\
n_{A0}+n_{B0}+n_{V0}&=N_0\label{lmconstraint3}\\
n_{A1}+n_{B1}+n_{V1}&=N_1\label{lmconstraint4}\\
n_{V0}&=0\label{lmconstraint5}
\end{align}
where all quantities on the right-hand side are considered to be
independant of $n_{B0}$.
The first two constraints follow from conservation of atom number
by component; the second two from conservation of site number;
and the last constraint neglects vacancies in the interface.
From these constraints, we can see that
$n_{A0}'=-1$, $n_{A1}'=1$, $n_{B1}'=-1$, $n_{V0}'=0$, and $n_{V1}'=0$.
Substituting these values into equation (\ref{generalisotherm})
and rearranging leads to equation (\ref{mcleanisotherm}),
after identifying 
$x_A=n_{A1}/(n_{A1}+n_{B1})$,
$x_B=n_{B1}/(n_{A1}+n_{B1})$,
$\Gamma_B/\Gamma_0=n_{B0}/N_0=x_{B0}$,
and $x_{A0}=1-x_{B0}$.

A similar result may be obtained by
replacing the final constraint (\ref{lmconstraint5}) 
with the equation
$n_{A1}+n_{B1}=n_1$, which
permits the interface and the bulk
to exchange an atom of component A
for an atom of component B while allowing no change in
the total number $n_1$ of atoms in the bulk.
This relation leads to the same set 
of exponents as in the previous case, but
the interface vacancy content
is no longer necessarily zero. 
As a consequence
we must write
$x_{B0} = (1+x_{V0})\Gamma_B/\Gamma_0$ and
equation (\ref{mcleanisotherm}) instead reads
\begin{align}
\frac{\Gamma_B}
{\overset{~}{\Gamma_V}-\Gamma_B}
=\frac{x_{B1}}{x_{A1}}
\text{e}^{-\delta G/k_BT}
\label{mcleanisovacs}
\end{align} 
with ${\Gamma_V}\equiv\Gamma_0/(1+x_{V0})$.
This constraint accounts for interface vacancies
by normalizing the 
maximum segregation to match a given 
interface vacancy content. 
In either case, the nature of the final constraint
indicates that this relation best models segregation
in a system that does not allow
variable vacancy content, whether 
in the interface or in the bulk.

The Langmuir-McLean isotherm therefore
represents the most probable
segregation for a given bulk atom density $n_1$.
Let us instead seek the most probable segregation
for a specified bulk impurity composition $x_{B1}$.
To do so, consider replacing constraint
(\ref{lmconstraint5}) with the equation $n_{B1}/n_{A1}=r$,
where $r$ is a fixed positive number.
Fixing $r$ also fixes $x_{B1}$ because $x_{B1}=r/(1+r)$.
It can be shown that in this case we have
$n_{A0}'=1/r$, $n_{A1}'=-1/r$, $n_{B1}'=-1$, $n_{V0}'=-(1+r)/r$, 
and $n_{V1}'=(1+r)/r$, leading to
\begin{align}
n_{B0}n_{A0}^{1/r}n_{V0}^{-(1+r)/r}=
n_{B1}n_{A1}^{1/r}n_{V1}^{-(1+r)/r}
\text{e}^{-\delta G/k_BT}
\end{align}
or in terms of mole fractions
\begin{align}
\frac{x_{B0}}{x_{A0}}
\left(\frac{x_{A0}}{x_{V0}}\right)^{1/x_{B1}}=
\frac{x_{B1}}{x_{A1}}
\left(\frac{x_{A1}}{x_{V1}}\right)^{1/x_{B1}}
\text{e}^{-\delta G/k_BT}
\label{molfracisotherm}
\end{align}
where 
$x_{V0}=n_{V0}/(n_{A0}+n_{B0})$
and 
$x_{V1}=n_{V1}/(n_{A1}+n_{B1})$.
This system is constrained such that
if the bulk loses
a single solute atom to the interface, 
it must also lose $1/r$ solvent atoms
to maintain a constant composition,
and so gain $(1+r)/r$ vacancies.

Equation (\ref{molfracisotherm}) is the 
central focus of this study.
In the following I present numerical solutions
and discuss some of its implications.

\section{Numerical analysis}

To investigate the extent to which the solutions to 
(\ref{mcleanisotherm}) and
(\ref{molfracisotherm}) differ, I have 
determined interface compositions $x_{B0}$ that 
satisfy equation (\ref{molfracisotherm})
as a function of bulk composition $x_{B1}$
for specific values of $\nu=x_{V0}/x_{V1}$
and $\delta G/k_BT$ using numerical techniques.
Explicitly, I have defined
\begin{align}
f(x,y)=
& y(1-y)^{(1-x)/x}v_0^{-1/x}\nonumber\\
&-x(1-x)^{(1-x)/x}v_1^{-1/x}
e^{-\delta G/k_BT}
\label{interp}
\end{align}
and interpolated to find the set of points $(x_0,y_0)$
such that $f(x_0,y_0)=0$. Given $v_0=x_{V0}$,
$v_1=x_{V1}$, and $\delta G$, the points $(x_0,y_0)$ 
represent solutions to equation (\ref{molfracisotherm}),
with $y_0=x_{B0}=\Gamma_B/\Gamma_V$ for a specified
bulk composition $x_0=x_{B1}$.

\begin{figure}[H]
\center
\includegraphics[width=\textwidth]{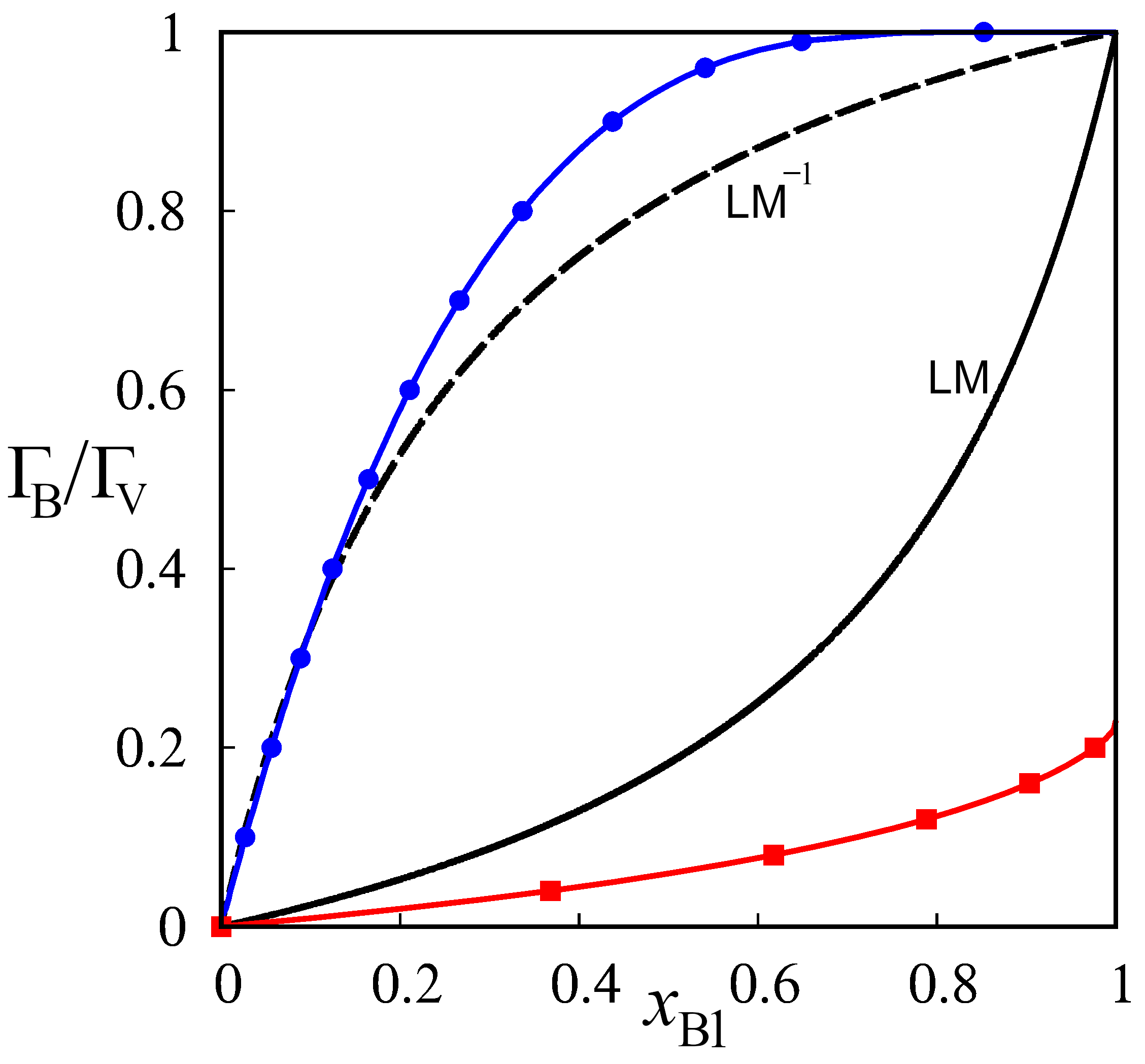}
\caption{The segregation ratio
$x_{B0}=\Gamma_B/\Gamma_V$ as predicted by equation
(\ref{molfracisotherm}) for segregation free energy
$\delta G=+1.5k_BT$, 
plotted versus mole fraction of bulk solute
$x_{B1}$, with equal vacancy mole fractions
in the bulk and in the interface ($x_{V0}=x_{V1}$).
The Langmuir-McLean isotherms for $\delta G=+1.5k_BT$
and $\delta G=-1.5k_BT$ are labeled $LM$ and $LM^{-1}$,
respectively.
All curves are vacancy-normalized, 
with $\Gamma_V=\Gamma_0/(1+x_{V0})$.
Both upper and lower curves are solutions to
equation (\ref{molfracisotherm}) for $\delta G=+1.5k_BT$.
With each value $x_{B1}$ is associated two
possible stable compositions, or complexions: 
one on the blue curve and one on the red curve.
}
\label{figure1}
\end{figure}

A typical solution set is plotted in Figure (\ref{figure1}),
for which $\delta G/k_BT=+1.5$ and $\nu=1.0$.
In the same figure I have plotted the 
associated Langmuir-McLean isotherm, labeled
$LM$, as well as the Langmuir-McLean isotherm
for $\delta G/k_BT=-1.5$, labeled $LM^{-1}$. 
It can be seen
that across a wide range of compositions
the lower curve predicts segregation at rates
lower than those of the Langmuir-McLean isotherm.
The most striking difference, however, is 
the appearance of a second branch,
as indicated by the blue curve with round bullets. 
In this figure the second branch 
tracks $LM^{-1}$ in the dilute limit.
Unlike $LM$ and $LM^{-1}$, which correspond
to oppositely signed segregation free energies,
both red and blue curves represent solutions
to equation (\ref{molfracisotherm}) for 
$\delta G/k_BT=+1.5$.
The red curve with square bullets
represents a solution with a
diminished concentration
of solute in the interface than in the bulk
($x_{B0}<x_{B1}$)
whereas the blue curve with round bullets 
represents a solution
with augmented
solute content in the interface
($x_{B0}>x_{B1}$).

\begin{figure}[H]
\center
\includegraphics[width=0.95\textwidth]{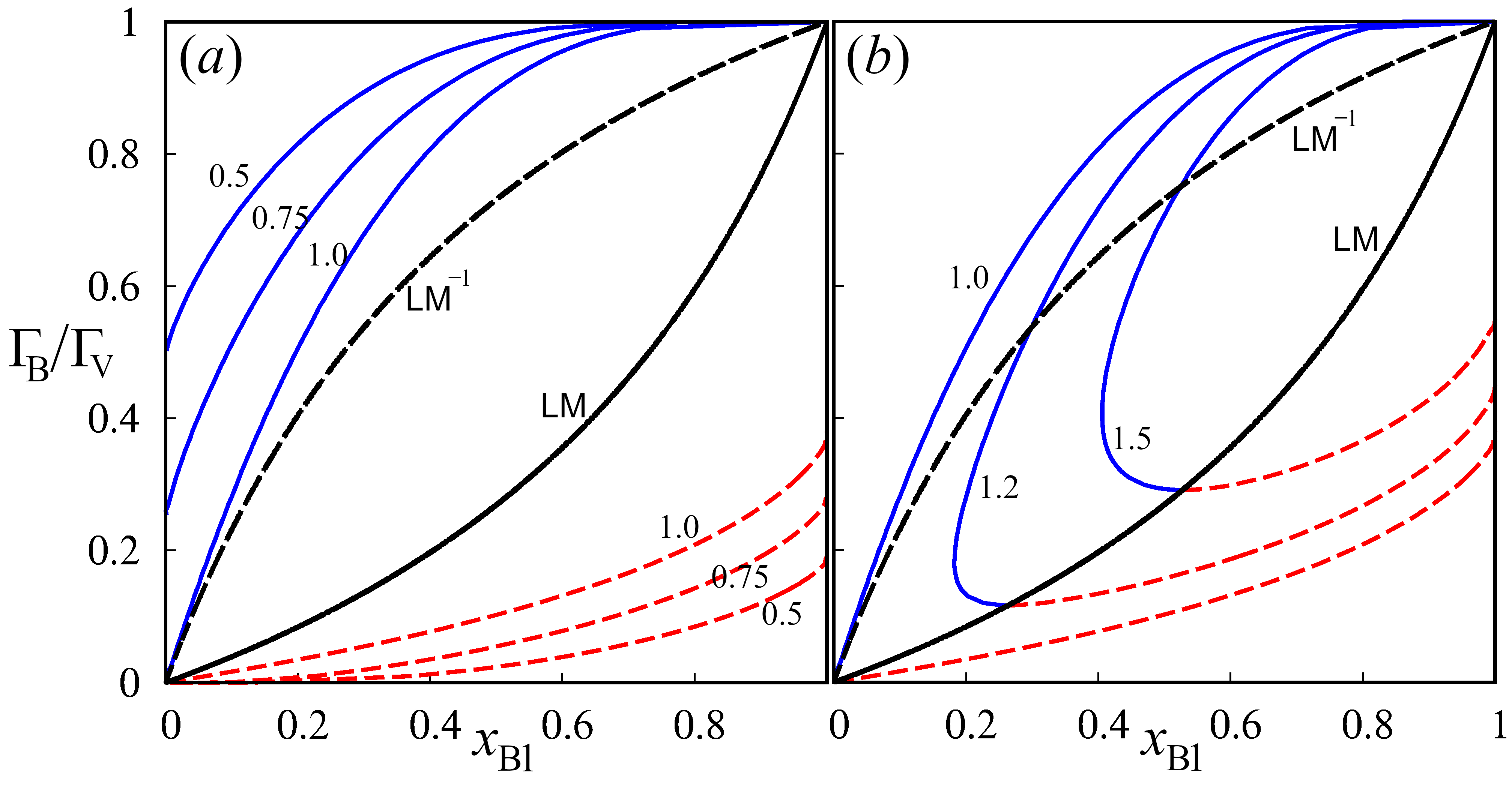}
\caption{\textit{Dependance on vacancy content ratio}.
The segregation ratio
$x_{B0}=\Gamma_B/\Gamma_V$ as predicted by equation
(\ref{molfracisotherm}), plotted
for several values of 
$\nu=x_{V0}/x_{V1}$.
All curves in both (a) and (b) 
represent grain boundaries with
segregation free energy $\delta G=k_BT$
and are labeled with an associated
value of $\nu$.
The curve labeled $LM$ is the
Langmuir-McLean isotherm for $\delta G=+k_BT$,
and the curve labeled $LM^{-1}$
is the Langmuir-McLean isotherm for $\delta G=-k_BT$.
(a) $\nu=1.0$, $\nu=0.75$, and $\nu=0.5$. Each value
admits an upper and a lower branch.
(b) $\nu=1.0$, $\nu=1.2$, and $\nu=1.5$. The
two branches merge and pull away from (0,0) for $\nu>1$.
}
\label{figure2}
\end{figure}

The nature of the isotherm fundamentally alters
if the vacancy ratio $\nu$ differs from unity.
I illustrate the dependance on $\nu$ 
in Figure (\ref{figure2}),
where solutions obtained for 
the same segregation free energy 
$\delta G/k_BT=+1.0$
but differing values of $\nu$ are plotted.
Figure (\ref{figure2}a) depicts variations that occur
for $\nu\leq 1$, when the mole fraction of
vacancies in the bulk exceeds that in the interface. 
As $\nu$ decreases, it can be seen that a gap opens in
the upper branch (blue)
along the $x_{B0}$ axis, suggesting a
minimal segregation $x_{B0}\approx 1-\nu$, 
i.e. $\Gamma_B\approx (v_1-v_0)\Gamma_0/(v_1+v_0)$,
in the dilute limit.
The lower branch shifts uniformly downward as $\nu$
increases, indicating
reduced segregation roughly proportional to $\nu$.

A system for which the interface vacancy content exceeds
the bulk vacancy content ($\nu >1)$ is depicted in
Figure (\ref{figure2}b). It can be seen that the upper 
and lower branches pull away from the origin and merge
as $\nu$ increases, opening up a gap along the $x_{B1}$ 
axis in which no potential stable solutions $x_{B0}$ exist,
apart from $x_{B0}=0$ or $x_{B1}=1$. 
Calculations suggest that this gap exists
even for small excursions in $\nu$ above 1.

\begin{figure}[H]
\center
\includegraphics[width=0.95\textwidth]{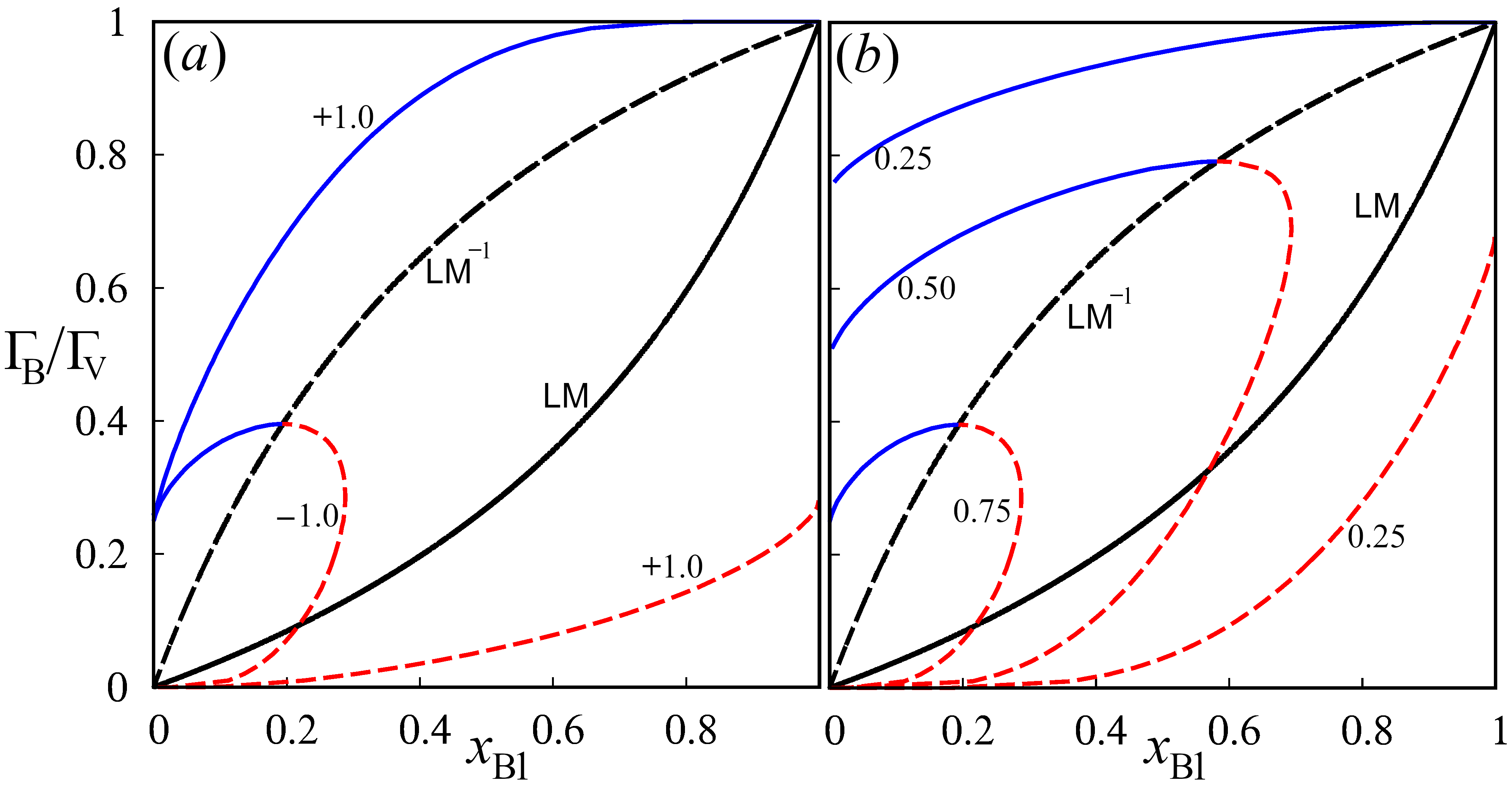}
\caption{\textit{Negative segregation free energy.}
The segregation ratio
$x_{B0}=\Gamma_B/\Gamma_V$ as predicted by equation
(\ref{molfracisotherm}) plotted for several values of
$\delta G<0$.
The curve labeled $LM$ is the
Langmuir-McLean isotherm. 
The curve labeled $LM^{-1}$
is the Langmuir-McLean isotherm for $\delta G=-k_BT$.
(a) Comparison between solutions obtained for 
$\delta G=+k_BT$ and
$\delta G=-k_BT$. Both solutions have been 
obtained assuming $\nu=0.75$. 
In the negative energy case the upper and lower branches
have merged to form the small inner half-loop about the
origin.
(b)
Three curves with $\delta G=-k_BT$ but different
values of $\nu$. As $\nu$ increases the upper
and lower branches pull away from the $(1,1)$
and merge.
The curves vanish for $\nu\geq 1$.
}
\label{figure3}
\end{figure}

The solutions to equation (\ref{molfracisotherm}) 
exhibit much different behavior than the Langmuir-McLean
isotherm under segregation free energy sign reversal.
In Figure (\ref{figure3}a) I depict two curves
obtained for $\delta G/k_BT=+1.0$ and $\delta G/k_BT=-1.0$,
both using $\nu=0.75$. Also depicted are the 
associated Langmuir-McLean isotherms. Whereas the
Langmuir-McLean isotherm flips across the diagonal,
solutions to equation (\ref{molfracisotherm}) assume
very different forms. Note that the two upper branches
yield the same minimal segregation 
$x_{B0}\approx 1-\nu$ as $x_{B1}\rightarrow 0$.

The dependance on $\nu$ for negative segregation free
energy solutions is explored in Figure (\ref{figure3}b).
All of the curves plotted in Figure (\ref{figure3}b) were
obtained using $\delta G/k_BT=-1.0$, except for the
curve $LM$, which is the Langmuir-McLean isotherm
for $\delta G/k_BT=+1.0$. The curves are labeled with
the associated value of $\nu$. As $\nu$ increases
toward 1, the isotherm pulls away from the point $(1,1)$
and contracts toward the origin. For $\nu=1$ the isotherm
vanishes. No solutions exist for $\delta G<0$ and $\nu\geq 1$,
apart from $x_{B0}=0$ or $x_{B0}=1$.

\section{Discussion}

To more readily compare equations 
(\ref{mcleanisotherm}) and (\ref{molfracisotherm}),
note that we can express (\ref{molfracisotherm}) as
\begin{align}
\frac{\Gamma_B}{\Gamma_V-\Gamma_B}=
\frac{x_B}{x_A}
\text{exp}\left[
-\frac{\delta G}{k_BT}
+\frac{1}{x_{B}}\ln
\left(\frac{x_{A}x_{V0}}{x_{A0}x_{V1}}\right)
\right]
\end{align}
with $\Gamma_V=\Gamma_0/(1+x_{V0})$ as before.
It can be seen from this equation
that if the composition and vacancy content
are similar in the interface and the bulk, 
then the ratio inside the logarithm is close to unity,
reproducing the original Langmuir-McLean relation.
This ratio may be recognized
as an approximation to the equilibrium rate 
constant $k$ for the interaction
\begin{align}
A_0+V_1\overset{k}{\leftrightharpoons} A_1+V_0
\end{align}
in which the interface and the bulk exchange
component A and vacancies.
We would therefore expect $k\approx 1$ when the
standard formation energies for vacancies and
component A do not differ much between bulk and
interface sites. Otherwise, unconditional reduction to
the Langmuir-McLean form
requires $r=-1$, which is unphysical.

When $k\neq 1$, the numerical
analysis presented in the previous section indicates
qualitative differences between segregation described
by the Langmuir-McLean model
and by equation (\ref{molfracisotherm}). 
The appearance of two stable states, or
complexions, over a broad range of values 
indicates that there are two different
system configurations that can accomodate
chemical differences between the interface
and the surrounding bulk. 
The nature of these two configurations
is unclear, apart from the fact that
one is enriched, and one depleted, in segregated solute,
relative to the bulk.
From a purely mathematical perspective,
these configurations result from the fact that 
$x(1-x)^{1/r}=C$
can admit two distinct solutions, where $r$ and $C$ are constants.

The appearance of two branches in solutions
to equation (\ref{molfracisotherm}) 
is not uncommon. 
Each branch indicates a set of points
such that $F'=0$.
The stability of each state can be determined
by evaluating the second derivative $F''$.
In the Langmuir-McLean approximation, with $U''=0$,
we obtain
\begin{align}
\frac{F''}{k_BT}=
& \frac{1}{n_{B0}}
+\frac{1}{n_{B1}}
+\frac{1}{r^2}\left[\frac{1}{n_{A0}}
+\frac{1}{n_{A1}}\right]\nonumber\\
&+\frac{(1+r)^2}{r^2}\left[\frac{1}{n_{V0}}
+\frac{1}{n_{V1}}\right]
\label{lap}
\end{align}
from which it follows that both branches represent
potentially stable solutions as long as the 
$\lbrace n_{\alpha}\rbrace$ are
all positive. In a more realistic model,
$U''<0$ could potentially modify the stability 
of either branch.

We might instead expect
one stable branch and one unstable branch,
and so it is important to question whether
both branches are physically relevant.
Due to the introduction of mole fractions,
no mechanism exists in the formalism to guarantee that
all component population variables $\{n_{\alpha}\}$ 
remain individually 
positive in the solution.
Indeed, negative population variables easily appear as solutions
to the traditional Langmuir-McLean equation (\ref{mcleanisotherm})
for most values of $x_B$
once concrete values are specified for the model parameters.
In the current model, 
negative values could lead to $F''<0$
in equation (\ref{lap}), resulting in an unphysical
solution that appears to be thermodynamically viable.

Therefore let us investigate whether either solution
requires negative population variables.
At every point along either branch it is clear that
$0<x_{B0}<1$ and $0<x_{A0}<1$, so that
$n_{B0}$ and $n_{A0}$ must be either 
both positive or both negative.
Also, $n_{V0}$ must have the same sign as 
$n_{B0}$ and $n_{A0}$ when we provide an appropriate value 
for $x_{V0}$ to define the quantity
$\Gamma_V=\Gamma_0/(1+x_{V0})$.
The same considerations apply for $n_{A1}$, $n_{B1}$, 
and $n_{V1}$, except that the sign of $n_{V1}$
is linked to the sign of $n_{V0}$
through the quantity $\nu=x_{V0}/x_{V1}$;
only positive values for $\nu$ have been considered
in this work. 
These considerations suggest that 
all quantities in the solution
are either all positive or all negative.
But it is clear that equation (\ref{molfracisotherm})
is invariant under a transformation that inverts
the sign of all population variables.
If $\{n_{\alpha}\}$ is a solution, so is $\{-n_{\alpha}\}$.
The corresponding mole fractions
are positive in both cases and satisfy the same equation.

Each of these branches therefore
represents a set of potentially stable, physically 
relevant solutions to equation (\ref{molfracisotherm}).
As in the Langmuir-McLean case, however, the
entire domain is most likely not accessible
once concrete parameters have been specified.
It seems probable that when the system
finds itself in one of these two states, the second
state becomes both unphysical and unstable, 
corresponding to negative $\{n_{\alpha}\}$ and 
$F'' < 0$. 

The free energy $F$ 
and its derivatives $F'$ and $F''$
in the Stirling approximation
become difficult to define along the borders,
where at least one population variable equals zero.
The nature of the limiting behavior of the system 
at the poles (0,0) and (1,1) clearly
influences the shape of the global isotherm.
The value of $\nu$ and the sign of $\delta G$ appear
to control whether the system is attracted or repulsed
from these poles, and to what extent. This suggests
that the local value of $\nu$ plays a large role
in controlling the dynamics of segregation.

This model may be most appropriate in systems that
exhibit a preferred bulk solute content $x_{B1}$. 
On the other hand, the constant mole fraction constraint
on which it is based is better aligned with 
the interpretation that it provides the most probable
segregation for a specified bulk composition.
Regardless of its applicability, the substantial departure 
observed from Langmuir-McLean behavior indicates
the critical role that the vacancy constraint plays 
in determining the shape of the Langmuir-McLean isotherm.

\section{Summary}
In this work I have presented 
a simple modification to the
Langmuir-McLean model of grain boundary segregation,
leading to equation (\ref{molfracisotherm}).
In contrast to the Langmuir-McLean model, 
the proposed model allows the interface
and the bulk to exchange vacancies as well as atoms 
to determine the most probable
segregation given a specified
bulk mole solute content $x_{B1}$.

Numerical analysis indicates that
this modification has a large effect
on the predicted segregation.
In particular, two complexions 
appear across a wide range of bulk compositions,
corresponding to solute enrichment or deficiency 
relative to the bulk.
The ratio of the vacancy
mole fraction in the interface
to the vacancy mole fraction in the bulk
assumes a prominent role
in determining the shape of the isotherm.
\end{multicols}

\section*{Acknowledgements}
This research did not receive any specific grant from
funding agencies in the public, commercial, or not-for-profit sectors.

\section*{References}
\citations


\end{document}